\documentclass[12pt,preprint]{aastex} 
\usepackage{natbib}
\newcommand{\degree}{\hbox{$^\circ$}}
\newcommand{\etal}{et\,al.}
\newcommand{\fuse}{{\it FUSE}}
\newcommand{\halpha}{H$\alpha$}
\newcommand{\gsim}{\raise0.3ex\hbox{$>$}\kern-0.75em{\lower0.65ex\hbox{$\sim$}}}
\newcommand{\kms}{km\,s$^{-1}$}
\newcommand{\lsim}{\raise0.3ex\hbox{$<$}\kern-0.75em{\lower0.65ex\hbox{$\sim$}}}
\newcommand{\msun}{M$_{\odot}$}
\newcommand{\osix}{\ion{O}{6}}
\newcommand{\OVI}{O~{\sc vi}}
\newcommand{\HI}{H~{\sc i}}
\newcommand{\HII}{H~{\sc ii}}

\newcommand{\CII}{C~{\sc ii}}
\newcommand{\NI}{N~{\sc i}}

\newcommand{\OI}{O~{\sc i}}
\newcommand{\SiII}{Si~{\sc ii}}
\newcommand{\SIII}{S~{\sc iii}}
\newcommand{\PII}{P~{\sc ii}}
\newcommand{\ArI}{Ar~{\sc i}}
\newcommand{\FeII}{Fe~{\sc ii}}

\begin{document}     
\slugcomment{Astrophysical Journal, in press}
%-----------------------------------------------------------------------------%
\title{Probing The Multiphase Interstellar Medium Of The Dwarf Starburst 
Galaxy NGC 625 With {\it FUSE} Spectroscopy\footnote{Based on observations made
with the NASA-CNES-CSA {\it Far Ultraviolet Spectroscopic Explorer. FUSE} is 
operated for NASA by the Johns Hopkins University under NASA contract NAS 
5-32985.}}
%-----------------------------------------------------------------------------%
\author{John M. Cannon}
\affil{Max-Planck-Institut f{\"u}r Astronomie, K{\"o}nigstuhl 17, D-69117, 
Heidelberg, Germany}
\email{cannon@mpia.de}

\author{Evan D. Skillman}
\affil{Department of Astronomy, University of Minnesota, 116 Church St. 
S.E., Minneapolis, MN 55455}
\email{skillman@astro.umn.edu}

\author{Kenneth R. Sembach}
\affil{Space Telescope Science Institute, 3700 San Martin Drive, 
Baltimore, MD 21218}
\email{sembach@stsci.edu}

\author{Dominik J. Bomans}
\affil{Astronomisches Institut, Ruhr-Universit{\" a}t Bochum, 
Universit{\" a}tsstr. 150, 44780, Bochum, Germany} 
\email{bomans@astro.ruhr-uni-bochum.de}
%-----------------------------------------------------------------------------%
\begin{abstract}
%-----------------------------------------------------------------------------%

We present new \fuse\ spectroscopy of the dwarf starburst galaxy NGC\,625.
These observations probe multiple phases of the interstellar medium, 
including the coronal, ionized, neutral and molecular gas.  This nearby (D = 
3.9\,$\pm$\,0.2 Mpc) system shows a clear detection of outflowing coronal gas 
as traced by \ion{O}{6} $\lambda$\,1032 \AA\ absorption.  The centroid of the 
\ion{O}{6} profile is blueshifted with respect to the galaxy systemic velocity 
by $\sim$ 30 \kms, suggesting a low-velocity outflow.  The implied \ion{O}{6} 
velocity extent is found to be 100 $\pm$ 20 \kms, which is fully consistent 
with the detected \HI\ outflow velocity found in radio synthesis observations.
We detect multiple lines of diffuse H$_2$ absorption from the ISM of NGC\,625;
this is one of only a few extragalactic systems with \fuse\ detections of 
H$_2$ lines in the Lyman and Werner bands.  We find a potential abundance 
offset between the neutral and nebular gas that exceeds the errors on the 
derived column densities.  Since such an offset has been found in multiple 
dwarf galaxies, we discuss the implications of a lower-metallicity halo 
surrounding the central star forming regions of dwarf galaxies. The apparent 
offset may be due to saturation of the observed \OI\ line, and higher 
S/N observations are required to resolve this issue. 

\end{abstract}						


%-----------------------------------------------------------------------------%
\section{Introduction}
\label{S1}
%-----------------------------------------------------------------------------%

Local starbursting dwarf galaxies offer a unique opportunity to study the
various phases of the interstellar medium (ISM) under extreme conditions and 
in unprecedented detail. Furthermore, the physical conditions in these targets
are likely very similar to those found in galaxies at high redshifts, making 
them crucial benchmarks against which observations of such targets must be 
compared.  With these observations, we capitalize on the high-resolution 
capabilities of the {\it Far Ultraviolet Spectroscopic Explorer} (\fuse; see
{Moos \etal\ 2000}\nocite{moos00}) to observe the interaction of the multiple 
phases of the ISM in NGC\,625, a nearby \citep[D = 3.9\,$\pm$\,0.2 
Mpc;][]{cannon03}, low mass (M$_{\rm HI}$ = 1.1$\times$10$^8$ \msun; 
{Cannon \etal\ 2004}\nocite{cannon04a}), low metallicity (12 + log(O/H) $=$ 
8.14\,$\pm$\,0.02; {Skillman \etal\ 2003b}\nocite{skillman03b}) dwarf irregular
galaxy in the Sculptor Group that is currently undergoing a massive star 
formation episode (see Table~\ref{t1} for basic galaxy properties).

\placetable{t1}

Of paramount importance in the evolution of dwarf galaxies is the role played 
by outflows powered by the energy inputs of supernovae and stellar winds.  
These outflows may drive significant fractions of the gas from galaxies with 
shallow potential wells.  Recent simulations suggest that the return of metals
into the intergalactic medium (IGM) finds its most effective avenue in these 
low mass galaxies \citep{maclow99}.  Understanding the nature of these 
outflows is therefore a fundamental astrophysical problem with far-reaching 
implications.  These \fuse\ spectra isolate the powerful diagnostic absorption
lines of \osix\ $\lambda\lambda 1032, 1038$ \AA\ which probe the temperature 
and kinematic structure of the coronal gas in a galaxy \citep[see, 
e.g.,][]{heckman01a}.  The \osix\ ion traces hot ionized gas over the 
temperature range T $\sim$ 10$^5$ - 10$^6$ K, with maximum sensitivity 
to gas in collisional ionization equilibrium at temperatures T $\sim$ (2 - 
3)$\times$10$^5$ K \citep{sutherland93}. Thus, knowing the temperature of the 
emitting material, we can (with some assumptions) directly estimate electron 
density and pressure in the coronal gas.  We can then assess the importance 
and timescale of radiative cooling in outflows and in the starburst 
phenomenon.  Indeed, in a case such as NGC\,625 where the outflow is of 
comparatively low velocity (see, e.g., {Heckman \etal\ 
2001}\nocite{heckman01a}), such effects may be dominant. These observations 
thus provide an opportunity to investigate many properties of outflows in 
dwarf galaxies. 

Multiple independent lines of evidence suggest that the recent star formation
in NGC\,625 has been violent and that the coronal gas content is substantial. 
An extended soft x-ray component was detected in {\it ROSAT} imaging 
\citep{bomans98} and has been verified in newly-obtained {\it Chandra} imaging
(Cannon \etal, in preparation). To attain hot gas at large distances from the 
central starburst requires the presence of an outflow in the recent past. This 
outflow is also seen in \HI\ synthesis imaging \citep{cannon04a}, making 
NGC\,625 one of only a few dwarf galaxies with a detected neutral gas outflow. 
This outflow appears to be a result of the extended (both spatially and 
temporally) star formation event that the galaxy has undergone over the last 
\gsim\ 50 Myr (see {Cannon \etal\ 2003}\nocite{cannon03} for details).  This 
extended burst is unexpected, given the brevity of star formation implied by 
the presence of the $\lambda$\,4686 \AA\ spectroscopic Wolf-Rayet (W-R) 
feature (\lsim\ 6 Myr; see {Conti 1991}\nocite{conti91}; {Schaerer \etal\ 
1999}\nocite{schaerer99a}; the spectrum is presented in {Skillman \etal\ 
2003b}\nocite{skillman03b}). 

Theoretical and observational evidence suggests that star formation behavior 
should be a function of metallicity (see, e.g., {Maloney \& Black 
1988}\nocite{maloney88}; {Taylor \etal\ 1998}\nocite{taylor98}). NGC\,625 has 
a metallicity significantly below the solar value, [O/H] = $-$0.47\,$\pm$\,0.06
\citep{skillman03b}, which provides an opportunity to study the interplay 
between powerful starburst activity and molecular gas content at a metallicity
similar to that of the Small Magellanic Cloud.  Single-dish CO spectra reveal a
large CO cloud coincident with the main dust concentration (C{\^ o}t{\' e}, 
Braine, \& Cannon, in preparation), but offset from the current massive star 
formation regions.  This position also hosts a moderately-extincted, low-mass 
stellar cluster that produces a thermal radio continuum peak \citep{cannon04c}.

In light of these arguments, we also analyze the diffuse H$_2$ content of this
relatively low-metallicity galaxy.  The Lyman-Werner bands of H$_2$ are 
expected to be very sensitive to relatively cool diffuse H$_2$ gas (T $\sim$ 
100 - 1000 K) over a wide range of column densities (N(H$_2$) $\sim$ 10$^{14}$
- 10$^{22}$ cm$^{-2}$).  In observations of galaxies or large star-formation 
regions, these lines probe the diffuse H$_2$ content along many sightlines to 
ultraviolet (UV)-luminous sources.  Evidence suggests that diffuse H$_2$ clouds
are indeed prevalent at low metallicities; \citet{tumlinson02} detect diffuse 
H$_2$ along 92\%\ of sightlines toward UV sources in the SMC, which has very 
nearly the same metallicity \citep{dufour84,garnett99} as NGC\,625.  However, 
integrated \fuse\ spectra of metal-poor starbursts suggest that diffuse H$_2$ 
is more difficult to detect in distant targets ({Vidal-Madjar \etal\ 
2000}\nocite{vidalmadjar00}; {Heckman \etal\ 2001}\nocite{heckman01a}; {Thuan 
\etal\ 2002}\nocite{thuan02}; {Aloisi \etal\ 2003}\nocite{aloisi03}; 
{Lebouteiller \etal\ 2004}\nocite{lebouteiller04}; {Lecavelier des Etangs 
\etal\ 2004}\nocite{lecavelierdesetangs04}; {Hoopes \etal\ 
2004}\nocite{hoopes04}; see further discussion in \S~\ref{S4.2}).  With these 
data we add NGC\,625 to the small but growing sample of dwarf starburst systems
with \fuse\ observations of diffuse H$_2$ gas; we detect multiple H$_2$ lines 
and discuss the implied properties of the molecular gas.

Finally, since the spectral region probed by \fuse\ is rich in neutral and
ionized gas absorption lines, we also address the column densities and 
abundances of observable species in the ISM of NGC\,625.  This point is 
especially important since evidence is growing for a bimodal abundance 
distribution in the ISM of dwarf galaxies.  The nebular regions (i.e., near the
star formation regions and accessible to abundance studies via optical and 
near-infrared emission line spectroscopy) appear to have elevated abundances 
compared to the neutral interstellar gas through which \fuse\ sightlines 
usually pass.  The sample of dwarf systems with such an analysis is small, but 
the offsets appear to be pronounced ($>$ 0.5 dex in oxygen abundance).  With 
these arguments in mind, we analyze the neutral gas abundances in NGC\,625 and 
compare these to values obtained from nebular spectroscopy obtained by
\citet{skillman03b}.

Understanding the kinematics and behavior of the outflowing coronal gas, the 
molecular gas content, and the neutral-gas abundances are the major goals of 
this work.  With these data, we can characterize the interaction of 
the outflowing gas with the neutral components as revealed in our \HI\ imaging
and from other ionization species within the \fuse\ spectrum.  In addition, 
they allow us to estimate the metallicity of the neutral gas and compare
these values with those found for emission line gas in the nebular regions. 
We discuss the evolution of dwarf starburst galaxies using these \fuse\ data 
and published information in the literature.  

%-----------------------------------------------------------------------------%
\section{Observations and Data Reduction}
\label{S2}
%-----------------------------------------------------------------------------%

\begin{figure}
\epsscale{0.9}
\plotone{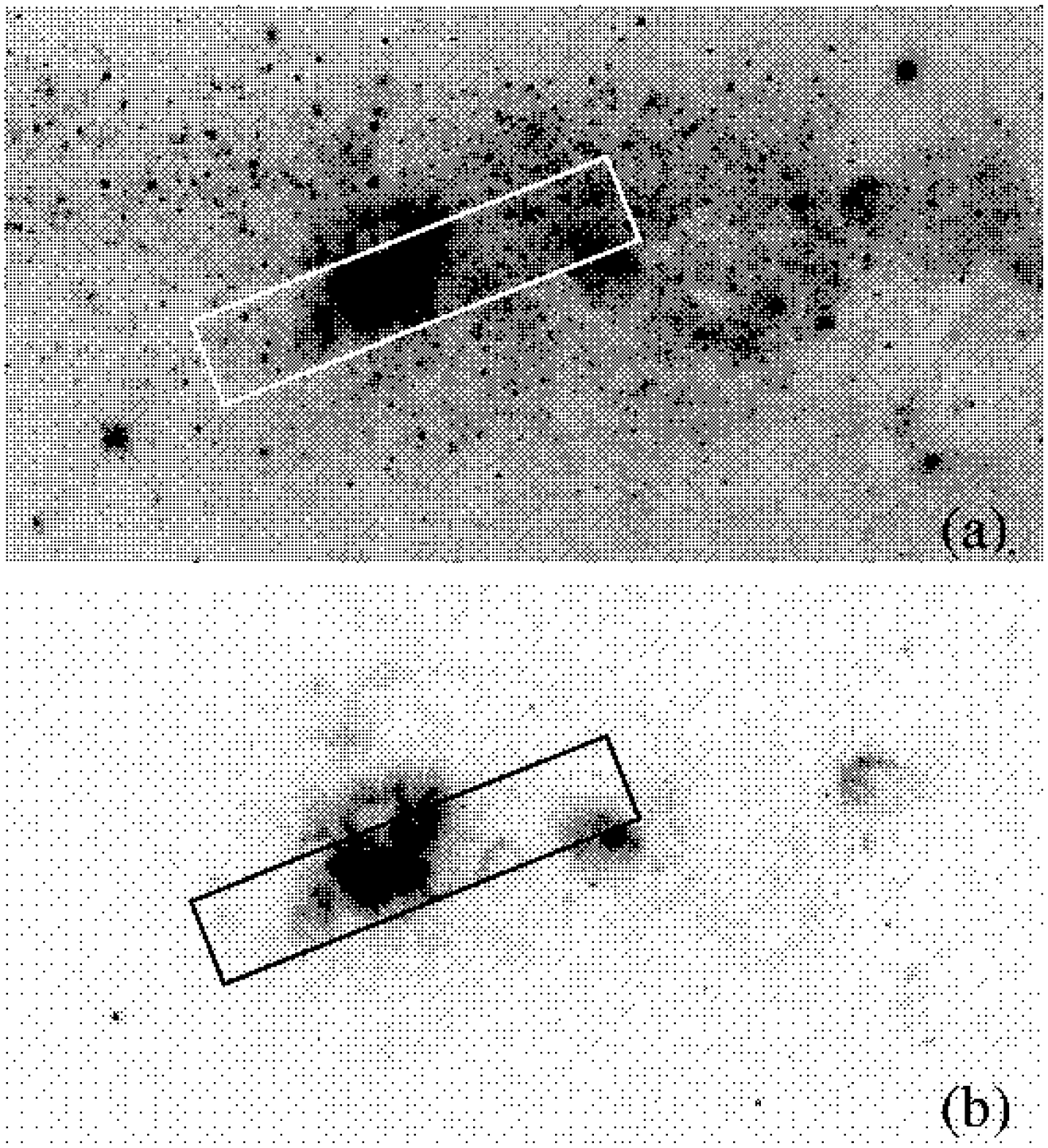}
\caption{Position of the LiF1 detector segment of the \fuse\ MDRS 
4\arcsec\,$\times$\,20\arcsec\ aperture overlaid on an {\it HST}/WFPC2 V-band 
image (a) and on an {\it HST}/WFPC2 continuum-subtracted \halpha\ image (b). 
The field of view is $\sim$ 875\,$\times$\,471 pc at the distance of 3.89 Mpc, 
with north up and east to the left. The average position angle during these 
observations was 112\degree\ east of north, as shown. We estimate the 
positional accuracy of the slit to be $\pm$\,1\arcsec, based on the nominal 
\fuse\ pointing accuracy ($\sim$ 0.5\arcsec\ rms), the {\it HST}/WFPC2 pointing
accuracy ($\sim$ 0.5\arcsec\ rms) and the comparison to ground-based 
astrometric solutions.  This slit placement is sensitive to the bulk of the UV 
flux from NGC\,625.}
\label{figcap1}
\end{figure}

\fuse\ spectroscopy of NGC\,625 using the 4\arcsec\,$\times$\,20\arcsec\ 
medium resolution aperture (MDRS) was obtained on 2003, November 8-9, for 
program D040. No roll angle constraints were placed on the observation, since
the aperture width is comparable to the size of the high-equivalent width 
\halpha\ emission (and hence the most UV-luminous sightlines).  Our 
observations were obtained with an average slit angle 112\degree\ east of 
north; the approximate aperture placement is shown superposed on {\it Hubble
Space Telescope} ({\it HST}) WFPC2 V-band and (continuum subtracted) \halpha\ 
images in Figure~\ref{figcap1}. The total integration time on-source was 58.1 
ksec, with $\sim$ 76\% of this occurring during orbital night, which decreases 
the contamination of the spectra by terrestrial \ion{N}{1} and \ion{O}{1} 
airglow lines.  Data from both orbital day and night were used in spectral 
regions not affected by airglow lines; otherwise, only data obtained during 
orbital night were used in our analysis of the NGC\,625 absorption features. 

This observation produced 23 raw time-tagged exposures which were re-processed
using a recent version of the \fuse\ calibration software (CALFUSE v2.4) 
available from Johns Hopkins University\footnote{See 
http://fuse.pha.jhu.edu/analysis/calfuse.html}.  This pipeline reduction 
removes mirror, grating, and spacecraft motions, and then corrects for 
astigmatism and Doppler motions.  Wavelengths are assigned to each photon, and
these events are screened for data quality (e.g., event bursts, spacecraft 
jitter, etc.).  The data are then flux calibrated and the bad pixel maps are 
corrected for spacecraft motions. Finally, spectral extractions are performed 
for each detector segment for each channel using either the orbital night-only 
data or the orbital day and night data.  Each individual exposure was processed
by the pipeline and the resulting spectra were combined to produce the final
spectrum for each channel and data segment. This results in a total of 16 
extracted spectra.

\begin{figure}
\plotone{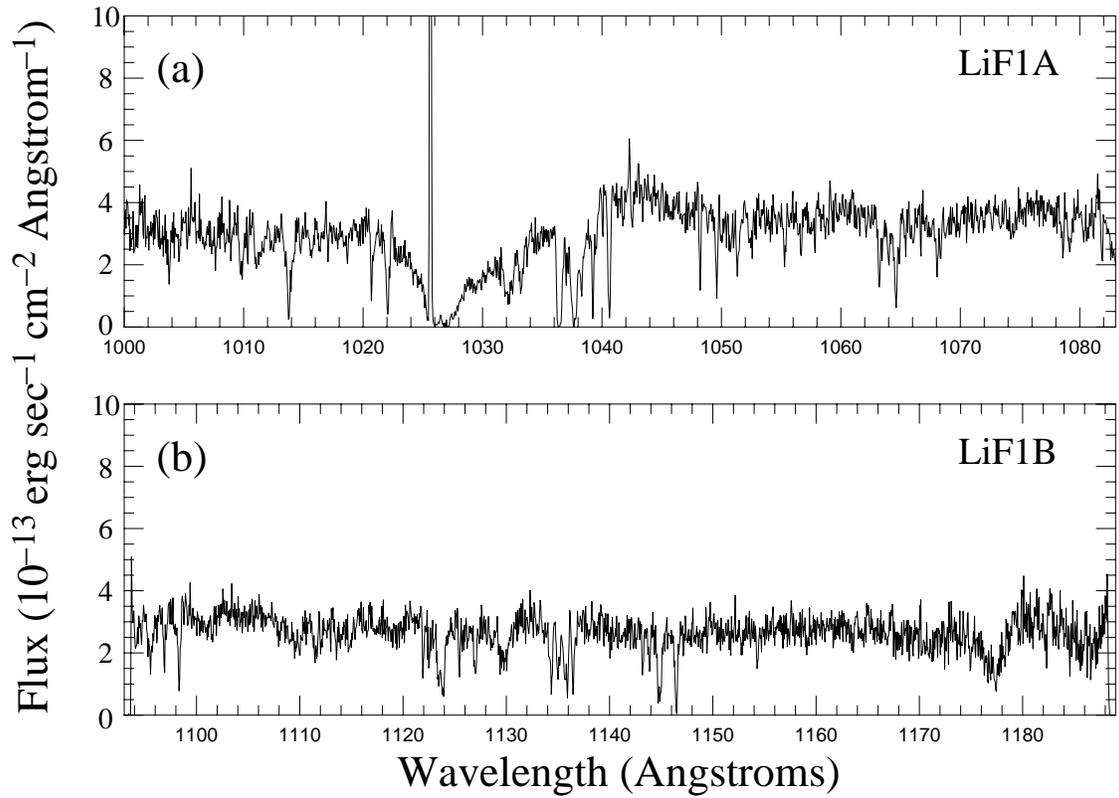}
\caption{Overview of our \fuse\ spectra, with the LiF1A spectrum shown in 
(a) and the LiF1B spectrum shown in (b).  These data have been binned by 8 
pixels ($\sim$ 0.05 \AA) for clarity.  Note the numerous absorption lines 
detected throughout this spectral region.}
\label{figcap2}
\end{figure}

\fuse\ uses four optical channels to produce 8 detector segments that allow 
spectral extractions between $\sim$ 900 and 1200 \AA. The two SiC channels are
optimized for shorter wavelengths (905 to 1100 \AA), while the LiF channels 
are optimized for longer wavelengths (980 to 1187 \AA).  The LiF channels are 
more sensitive than the SiC channels, so these provide our highest 
signal-to-noise (S/N) data.  The effective area of the telescope is also 
maximized at $\sim$ 1032 \AA\ (i.e., very near the important \osix\ absorption
features), so we concentrate our analysis on the LiF1 channel, with LiF2 
(lower S/N) serving to verify detections.  Data in the SiC channels are 
generally of low S/N (caused by a combination of lower effective area and 
channel misalignment between the LiF and SiC channels), and we do not use them 
in this analysis.  We present in Figure~\ref{figcap2} the LiF1A and LiF1B 
spectra of NGC\,625, binned to 0.05 \AA\ resolution for clarity. Note the 
richness of the spectrum, with numerous strong absorption lines detected. We do
not combine any of the overlapping detector segments, due to the changing 
spectral resolution and sensitivity of the detectors as a function of 
wavelength.  The final velocity resolution is $\sim$ 30 \kms, with S/N ratios 
of $\sim$ 10 and 5 per resolution element at 1032 \AA\ for the LiF1 and LiF2 
channels, respectively.  All velocities quoted are in the heliocentric 
reference frame. 

%-----------------------------------------------------------------------------%
\section{Analysis}
\label{S3}
%-----------------------------------------------------------------------------%

These \fuse\ data probe four different ISM phases: diffuse molecular gas, 
neutral gas, warm photoionized gas, and hot coronal gas.  We discuss the 
kinematics of each of these phases as derived from Gaussian fitting to the 
absorption profiles in the following subsection and summarize important line 
properties in Table~\ref{t2}.  We then discuss derived column densities of the
gas in \S~\ref{S3.2} and the inferred elemental abundances in \S~\ref{S3.3}.

%-----------------------------------------------------------------------------%
\subsection{Gas Kinematics}
\label{S3.1}
%-----------------------------------------------------------------------------%

There are two main absorbing components seen in these data.  First, Milky Way
gas appears near zero velocity.  Second, NGC\,625 absorptions are centered 
around $+$ 400 \kms.  In addition, there may also exist a halo 
intermediate-velocity cloud as seen in the red wings of some of the stronger 
Milky Way absorption lines (see below).  In this work we concentrate on the 
lines arising within NGC\,625.

Beginning with the molecular phase, we detect numerous low-level H$_2$ 
absorption lines in NGC\,625.  The Lyman and Werner bands of H$_2$ produce 
hundreds of absorption lines in the spectral region under study.  The lowest 
rotational levels (J = 0, 1, and 2) are expected to be most highly-populated by
the moderate diffuse molecular hydrogen columns and temperatures typically 
found in external galaxies (N(H$_2$) $\sim$ 10$^{14}$ - 10$^{16}$ cm$^{-2}$; 
T $\sim$ 100 - 1000 K). Of these low-level lines, six are detected in 
unblended, high S/N regions of the final spectra.  We do not detect any lines 
in the higher J levels. The velocity centroid of the detected lines (including
R(0)(2-0) $\lambda$\,1077.1399 \AA, R(1)(4-0) $\lambda$\,1049.960 \AA, 
R(2)(5-0) $\lambda$\,1038.689 \AA, P(1)(5-0) 1038.157 \AA, R(2)(0-0) 
$\lambda$\,1009.024 \AA, and R(1)(8-0) $\lambda$\,1002.449 \AA) is found to be
v$_{\rm H_2}$ = 401\,$\pm$\,10 \kms\ (all errors quoted in this work are 
1\,$\sigma$ unless otherwise noted).  This is in general agreement with, 
though slightly lower than, the \HI\ systemic velocity derived by 
\citet{cannon04a}, V$_{\rm sys}$ $=$ 413\,$\pm$\,5 \kms.  We note that our 
\fuse\ slit placement covers \HI\ gas in emission at velocities from $\sim$ 405
- 415 \kms.  Since velocity offsets of up to 20 \kms\ are common in \fuse\ 
spectra, we interpret the velocity centroid of the molecular gas to be 
coincident with the \HI\ and stellar populations.  We measure blueshifts of 
other species with respect to the velocity centroid for H$_2$ absorption, 
v$_{\rm H_2}$ = 401\,$\pm$\,5 \kms.

We detect absorption lines of N, O, Si, P, Ar and Fe that arise from the 
neutral gas phase (i.e., ionization potentials greater than or near that of H).
The velocity centroids of the strongest lines in clean spectral regions 
(including \ion{N}{1} $\lambda$\,1134.980 \AA, \ion{N}{1} 
$\lambda$\,1134.415 \AA, \ion{O}{1} $\lambda$\,1039.230 \AA, \ion{Si}{2} 
$\lambda$\,1020.699 \AA, \ion{P}{2} $\lambda$\,1152.818 \AA, \ion{Ar}{1} 
$\lambda$\,1048.220 \AA, \ion{Ar}{1} $\lambda$\,1066.660 \AA, \ion{Fe}{2}
$\lambda$\,1096.877 \AA, and \ion{Fe}{2} $\lambda$\,1063.176 \AA) yield an 
average neutral gas velocity of v$_{\rm neutral}$ = 392\,$\pm$\,10 \kms. This
can be compared with the molecular gas velocity derived above, v$_{\rm sys}$ 
$=$ 401\,$\pm$\,10 \kms.  These values are equal within the errors; the small
offset between the two may be caused by the neutral gas being split into two 
kinematic components (separated by 10-20 \kms).  To test for this, we compared
fits for single and double Gaussian components to the strongest line profiles 
(including \ion{O}{1} $\lambda$\,1039.230 \AA, \ion{Si}{2} 
$\lambda$\,1020.699 \AA, and \ion{Fe}{2} $\lambda$\,1063.176 \AA), but found no
statistically significant difference (in $\chi^2$ per degree of freedom) 
between them. The weakness of the putative second component introduces a 
negligible contribution to the error budgets for derived column densities and 
abundances (see below).

\begin{figure}
\plotone{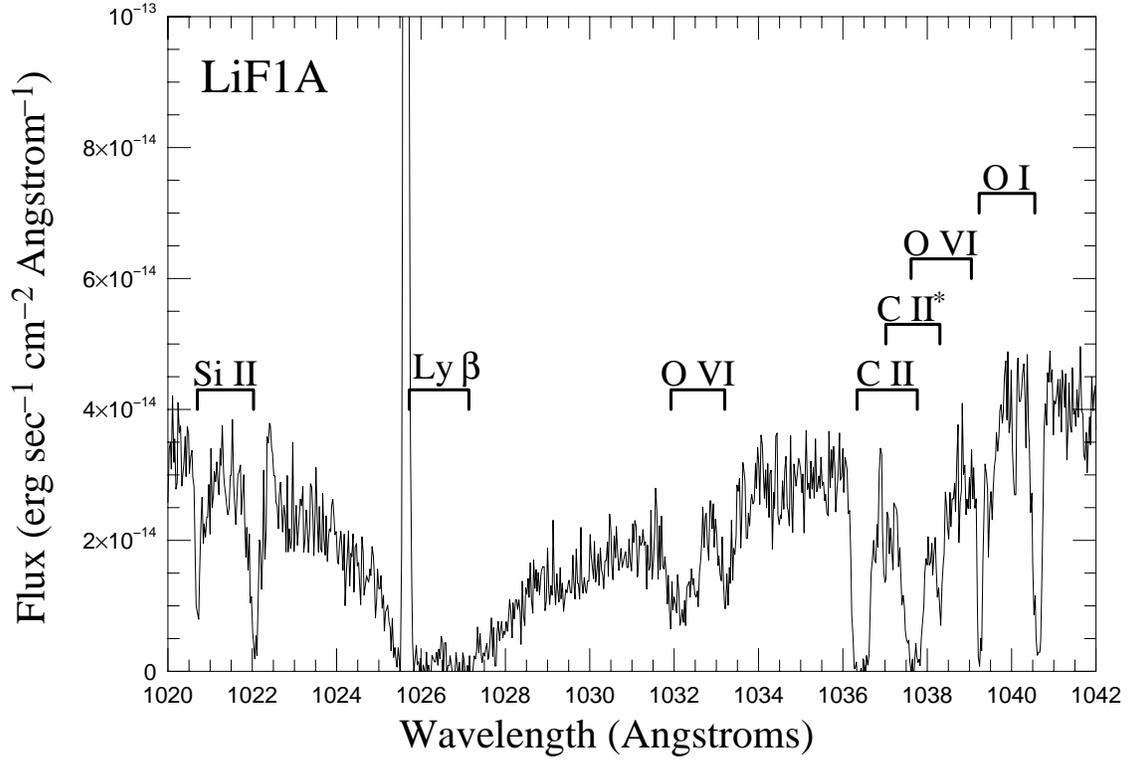}
\caption{Closer view of the spectral region surrounding Ly\,$\beta$ and the 
\osix\ absorption features, from the LiF1A detector segment. Prominent 
absorption lines are labeled with both the Milky Way and NGC\,625 (systemic) 
velocities.  The \osix\ absorption is strong (equivalent width $\sim$ 0.2 \AA) 
and the velocity centroid is blueshifted with respect to the molecular and 
neutral gas absorption centroids. See further discussion in \S~\ref{S3.1}, and 
also Figures~\ref{figcap4} and \ref{figcap5}.}
\label{figcap3}
\end{figure}

\begin{figure}
\plotone{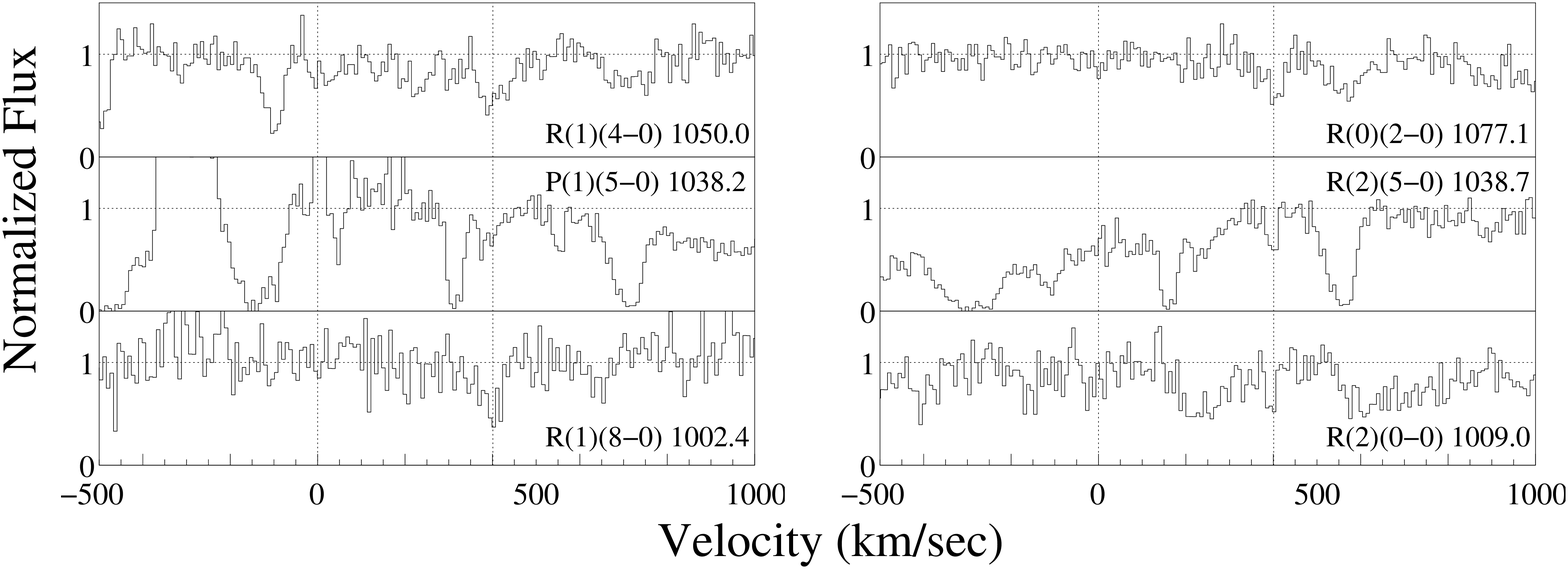}
\caption{Normalized line profiles of identified H$_2$ lines arising within the 
ISM of NGC\,625. Vertical lines show Milky Way and NGC\,625 velocities; these 
H$_2$ lines are discussed further in \S\S~\ref{S3.1} and \ref{S3.2}.}
\label{figcap4}
\end{figure}

\begin{figure}
\plotone{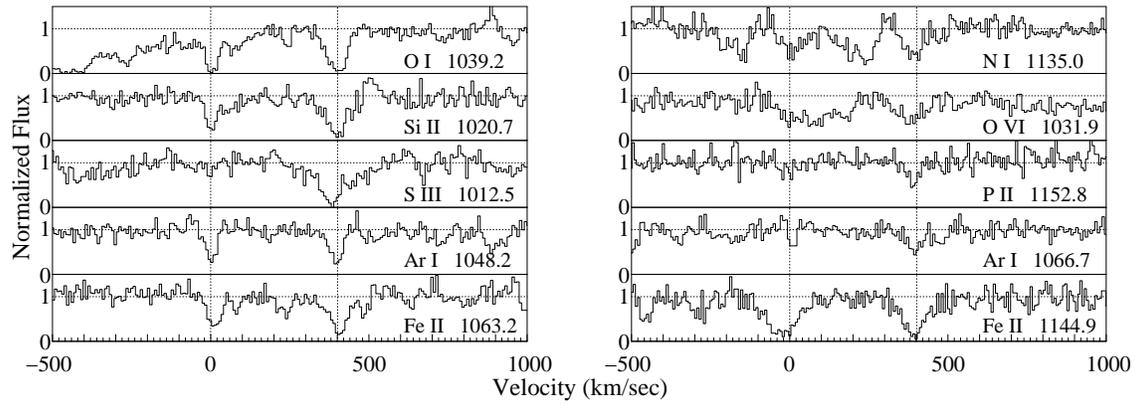}
\caption{Normalized line profiles of identified neutral and ionized atomic
absorption lines arising within the ISM of NGC\,625.  Vertical lines show 
Milky Way and NGC\,625 velocities; these absorption lines are discussed 
further in \S\S~\ref{S3.1} and \ref{S3.2}.}
\label{figcap5}
\end{figure}

There are two ionized gas absorption lines detected in clean regions of our 
spectra, including the $\lambda$\,1037.018 \AA\ line of \ion{C}{2}$^*$ and the 
$\lambda$\,1012.495 \AA\ line of \SIII.  These absorptions give an average 
ionized gas velocity of v$_{\rm ionized}$ = 377\,$\pm$\,9 \kms; i.e., offset 
with respect to the molecular gas by $-$24 \kms.  Finally, coronal gas is 
detected in \osix\ $\lambda$\,1031.926 \AA\ absorption at a velocity of 370 
\kms. This is blueshifted from the molecular gas velocity by $-$31 \kms, from 
the neutral gas by $-$22 \kms, and from the ionized gas by $-$7 \kms\ 
(though equal within errors).  The weaker line of the doublet at 
$\lambda$\,1037.617 \AA\ falls in a complicated spectral region and is not 
used in the present analysis; see Figure~\ref{figcap3} for a closer view of 
the spectral region around \osix. We present in Figures~\ref{figcap4} and 
\ref{figcap5} the normalized line profiles of the detected absorption lines 
from the ISM of NGC\,625 (absorption features were normalized by using clean 
spectral regions surrounding the lines; our error budget incorporates errors 
as a result of the continuum placements, which are usually comparable to the 
statistical noise). 

This scenario is consistent with a low-velocity outflow, where the coronal gas 
has the largest velocity offset from the starburst region (which presumably is 
coincident with the velocity of the molecular and neutral gas absorption 
features).  However, it is important to note that although the velocity 
centroids of the different types of gas may differ, there is considerable 
overlap in the velocity extents of many of the profiles (examine 
Figures~\ref{figcap4} and \ref{figcap5}).  We interpret these changing velocity
centroids and line breadths as results of the outflow, but it is clear that 
gas of all ionization levels co-exists throughout the galaxy.  This velocity 
gradient is discussed further in \S~\ref{S4.1}. 

%-----------------------------------------------------------------------------%
\subsection{Column Densities}
\label{S3.2}
%-----------------------------------------------------------------------------%

For each unblended spectral line shown in Table~\ref{t2}, the absorption 
profiles were converted into apparent optical depths, and then these profiles 
were integrated over velocity (see further discussion in {Savage \& Sembach
1991}\nocite{savage91}).  The essence of this technique is to measure the 
depth of normalized absorption lines as a function of velocity, and then to 
use atomic physics to infer the column density that best reproduces such a line
profile.  The column density and normalized line depth are related by:

\begin{equation}
N = \frac{m_e {c}}{{\pi}{e^2}{f}{\lambda}} \int ln\frac{I_0(v)}{I(v)} dv
\label{eq1}
\end{equation}

\noindent where N is the column density in atoms\,cm$^{-2}$, m$_e$ is the mass
of the electron, c is the speed of light, e$^2$ is the standard unit of charge,
{\it f} is the oscillator strength of the transition, $\lambda$ is the rest 
wavelength of the transition, and \begin{math}ln\frac{I_0(v)}{I(v)}\end{math} 
is the apparent optical depth as a function of velocity (the natural logarithm
of the ratio of intensity in the continuum versus intensity in the line, as a 
function of velocity). This method is preferred over other common practices 
(e.g., curve of growth analysis) because it does not presume a functional form
for the line under analysis.  In general, the results from standard curve of 
growth techniques and the apparent optical depth method agree very well, and 
we adopt the apparent optical depth method values in this work.  Note that the
kinematic characteristics were derived separately using Gaussian fitting (see 
above).

For the diffuse molecular hydrogen absorption detected, we derive weighted mean
column densities of log(N(H$_2$)) $=$ 14.92\,$\pm$\,0.19 (J\,$=$\,0), 
14.91\,$\pm$\,0.09 (J\,$=$\,1), and 14.47\,$\pm$\,0.13 (J\,$=$\,2).  Following
\citet{tumlinson02}, we can infer a range of rotational temperatures, T$_{01}$,
of $\sim$ 70-90 K.  At high densities where collisions dominate the level 
populations, this value is indicative of the kinetic temperature of the gas.
Such a progression is expected for relatively low-excitation H$_2$, with the 
bulk of the diffuse gas in the lowest available rotational levels, and the 
column densities rapidly decreasing toward higher rotational levels.  
Derivation of the \HI\ absorbing column is discussed in \S~\ref{S3.3}; 
combining this estimate with the measured molecular columns implies a diffuse 
molecular fraction {\it f}$_{\rm H_2} \sim$ 2$\times$10$^{-5}$.  It should be 
noted that this value samples the most UV-bright sightlines, and hence is 
heavily weighted toward material on the line of sight to the major 
starburst region.  This is not coincident with the highest \HI\ column density 
seen in emission (which is offset with respect to the starburst region; see 
{Cannon \etal\ 2004}\nocite{cannon04a}) nor with a large detected molecular 
cloud (from CO observations; see C{\^ o}t{\' e}, Braine, \& Cannon, in 
preparation).  In general this highlights a shortcoming of the Lyman and Werner
bands in probing cool molecular gas, as these lines are sensitive to diffuse 
gas (i.e., sightlines with low extinctions) but are less sensitive to the more 
common discrete molecular clouds and star formation complexes where the bulk of
the cool molecular material is expected to reside (since these regions will be 
more heavily extincted and hence have less background UV continuum emission). 
This point is discussed in more detail in \S~\ref{S4.2}.

For other neutral and ionized gas absorption lines present in our spectra, we 
derive the following column densities: log(N) $=$ 14.63 (\NI), 15.80 (\OI), 
15.37 (\SiII), 13.54 (\PII), 15.04 (\SIII), 13.98 (\ArI), and 14.73 (\FeII); 
typical errors are 0.1 - 0.2 dex (see Table~\ref{t2}).  We note that the \OI\  
$\lambda$\,1039.230 \AA\ line may be saturated; since the SiC channels are of
low S/N we cannot probe the amount of saturation empirically with other O lines
in these data alone. We discuss tests for saturation and their implications in 
\S~\ref{S3.3}.  For the purposes of this paper, we adopt the empirical \OI\ 
column (15.80\,$\pm$\,0.12) as a lower limit; we note that this implies an 
upper limit for our implied neutral gas [N/O] ratio (see next section). For 
the coronal gas as traced by \osix\ $\lambda$\,1031.926 \AA\ absorption, we 
find log(N) $=$ 14.32\,$\pm$\,0.08. We use the neutral species to derive 
gas-phase abundances in the next subsection; the coronal gas is discussed in 
more detail in  \S~\ref{S4.1}. 

\placetable{t2}

%-----------------------------------------------------------------------------%
\subsection{Relative Gas-Phase Abundances}
\label{S3.3}
%-----------------------------------------------------------------------------%

%-----------------------------------------------------------------------------%
\subsubsection{The Intrinsic \HI\ Column}
\label{S3.3.1}
%-----------------------------------------------------------------------------%

\begin{figure}
\plotone{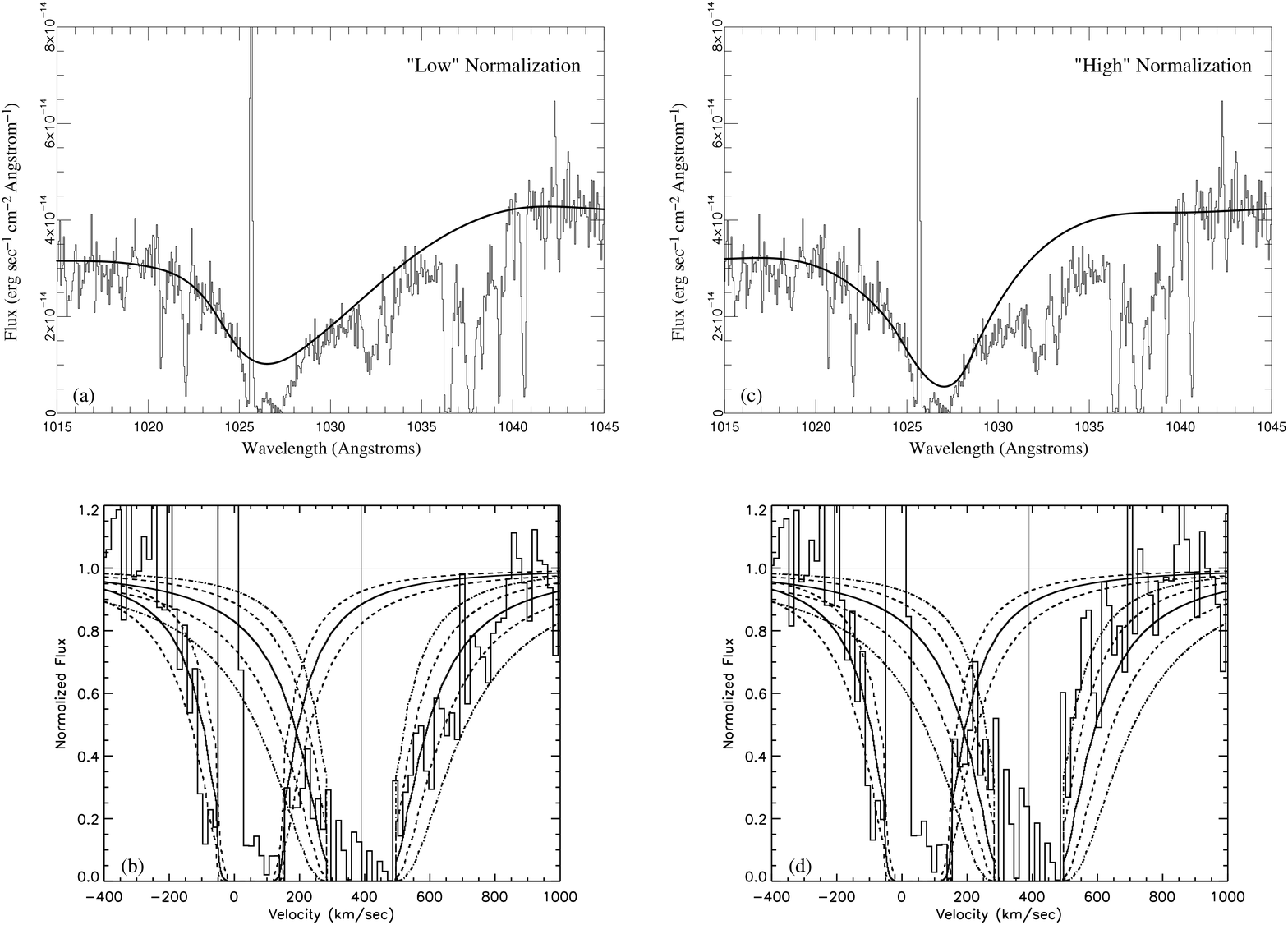}
\caption{Profiles of Galactic and NGC\,625 \HI\ Ly\,$\beta$ absorption.  (a)
and (c) show the ``low'' and ``high'' continuum normalizations, and (b) and (d)
show the resulting \HI\ absorption profiles.  Overlaid on (b) and (d) are 
profiles for the Galactic and NGC\,625 \HI\ Ly\,$\beta$ absorption.  For the 
Milky Way profile, three lines are shown at log(N(\HI)) $=$ 20.3 (solid line)
and log(N(\HI)) $=$ 20.1, 20.5 (dotted lines).  For NGC\,625, five lines are
shown at log(N(\HI)) $=$ 20.6 (solid), log(N(\HI)) $=$ 20.4, 20.8 (dotted 
lines), and log(N(\HI)) $=$ 20.2, 21.0 (dot-dash lines).  From these profiles
we deduce an absorbing \HI\ column of log(N(\HI)) $=$ 20.5\,$\pm$\,0.3 for 
NGC\,625, with the continuum normalization dominating the error budget.  See 
\S~\ref{S3.3} for further discussion.}
\label{figcap6}
\end{figure}

An estimate of the \HI\ column density in NGC\,625 from these data alone is 
hindered by the low S/N of the spectra in the SiC channels; this precludes us 
from using the higher-order Lyman lines, and we are left with only Ly\,$\beta$ 
to constrain the \HI\ column. However, the velocity separation of NGC\,625 and 
the Milky Way and the low measured Galactic column along this high-latitude 
sightline ($-$73.1\degree; we find a Galactic column log(N(\HI)) $=$ 
20.3\,$\pm$\,0.2) are sufficient to clearly separate Galactic and intrinsic 
absorption profiles.  This allows us to analyze both the red and blue wings of
the NGC\,625 \HI\ absorption feature, better constraining our fit to the \HI\ 
column density.  We present in Figure~\ref{figcap6} a closer view of the 
region surrounding Ly\,$\beta$, overlaid with fits to both the Galactic and 
intrinsic \HI\ columns.

A pronounced uncertainty remains in our measured \HI\ column due to the 
presence of strong stellar \osix\ P-Cygni profiles in the spectrum redward of
Ly\,$\beta$ (see Figure~\ref{figcap6}).  These features result in an 
uncertainty in the adopted continuum in the region around Ly\,$\beta$.  We 
adopt the mean of a ``low'' and ``high'' continuum placement, as shown in 
Figure~\ref{figcap6}.  This leads us to estimate the intrinsic \HI\ column in
NGC\,625 at log(N(\HI)) $=$ 20.5\,$\pm$\,0.3.  This is at least half a dex 
lower than the \HI\ 21\,cm emission column density in this region; comparing 
the \fuse\ aperture placement (see Figure~\ref{figcap1}) with the $\sim$ 
15\arcsec\ resolution \HI\ column density map presented by \citet{cannon04a}, 
we find that the \fuse\ aperture is completely enclosed within the 10$^{21}$
cm$^{-2}$ contour.  We believe that this estimate and associated uncertainty 
for the sight line probed are robust.  Using the 21\,cm value would result in 
a stronger Ly\,$\beta$ profile than the other models shown in 
Figure~\ref{figcap6} (see the dot-dash line corresponding to log(N(\HI)) $=$ 
21).  The difference between our absorption estimate and the 21\,cm value could
easily be due to the much different solid angles probed by the two 
measurements (a size ratio of 2.2:1 exists between the \HI\ emission beam and 
the solid angle of the {\it FUSE} MDRS aperture). We discuss the implications 
of this offset in \HI\ absorbing column in more detail in \S~\ref{S4}.

%-----------------------------------------------------------------------------%
\subsubsection{Neutral Gas Column Densities and Implied Abundances}
\label{S3.3.2}
%-----------------------------------------------------------------------------%

Using these column density estimates and assuming that the observed absorption 
lines arise from the main elemental species in the neutral ISM, we find the 
following abundances and elemental ratios: [N/H] $= -$1.80\,$\pm$\,0.34; [O/H]
$\ge -$1.36\,$\pm$\,0.33 (we stress again that the \OI\ $\lambda$\,1039.230 
\AA\ line may be saturated, and hence the lower limit is quoted; see 
further discussion in \S~\ref{S3.3.3}); [Ar/H] $= -$1.08\,$\pm$\,0.33; [Si/H] 
$= -$0.67\,$\pm$\,0.34; [Fe/H] $= -$1.22\,$\pm$\,0.34; log(N/O) $\le 
-$1.17\,$\pm$\,0.17 ([N/O] $\ge -$0.44\,$\pm$\,0.21).  These values use the 
updated solar abundances of {Holweger (2001}\nocite{holweger01}; N, Si, Fe) 
and {Asplund \etal\ (2004}\nocite{asplund04}; O), or the older standard values
of {Anders \& Grevesse (1989}\nocite{anders89}; Ar). These abundances are 
summarized in Table~\ref{t3}.  Note that a downward shift of the assumed \HI\ 
column leads to increased abundances, and that an increase in N(\HI) leads to 
a decrease in inferred metal abundance (specifically, if the \HI\ column is a 
factor of 2 larger, as might be inferred from the 21\,cm emission in this 
direction [see \S~\ref{S3.3.1}], the derived abundances will be a factor of 2 
lower than the quoted values).  Also, the N/O abundance ratio assumes
that the neutral gas components are co-spatial (i.e., that the ionization 
correction factor is negligible).

\placetable{t3}

Keeping in mind the relatively large errors involved with these absorption
line abundance measurements, and the points about potential line saturation
discussed in \S~\ref{S3.3.3}, these neutral gas values can be compared with 
the nebular abundances presented by \citet{skillman03b}.  Therein, for the 
major star formation region which dominates these \fuse\ sightlines, the 
calculated abundance ratios were found to be [O/H] $= -$0.47\,$\pm$\,0.06 and 
log(N/O) $= -$1.33\,$\pm$\,0.02.  Our neutral gas oxygen abundance (taken at 
face value) is found to be lower by $\sim$ 0.9 dex, while the relative 
abundances between N and O are very similar. The offset between the neutral 
and nebular gas abundances may indicate that the outer regions of dwarf 
galaxies are less enriched in heavy elements compared to the inner, 
star-forming regions.  We discuss this point further and compare to results 
found for other dwarf starbursts in the literature in \S~\ref{S4.3}.

It is interesting to note that \citet{skillman03b} found an elevated N/O ratio 
in the three highest surface brightness \HII\ regions in NGC\,625 when compared
to other star-forming dwarf galaxies.  This value was found to be comparable to
those for blue compact dwarfs, and was postulated to potentially arise as an 
effect of a long quiescent period prior to the current (temporally and 
spatially extended) star formation episode where the N abundance may be 
elevated by the evolution of intermediate mass stars.  Our neutral gas N/O
ratio (which is equal to the nebular value, within errors) is also found to be 
high compared to the neutral N/O values derived for other dwarf starbursts 
observed with \fuse\ (note that if the \OI\ $\lambda$\,1039.230 \AA\ line is 
saturated, the N/O value decreases, moving the NGC\,625 N/O ratio closer to  
the values found for other systems). While interpretation of the elevated N/O 
ratio in the neutral gas of starbursting dwarfs will require a larger 
statistical sample, it appears that in NGC\,625 the nucleosynthetic products 
of intermediate mass stars have been enriching the ISM for a long period, and 
that these products have been mixed into the neutral gas at large separations 
from the starburst regions (see also \S~\ref{S4.3}).

%-----------------------------------------------------------------------------%
\subsubsection{Effects of Potential Line Saturation}
\label{S3.3.3}
%-----------------------------------------------------------------------------%

As mentioned previously, the \OI\ $\lambda$\,1039.230 \AA\ line may be 
saturated, potentially severely underestimating the true \OI\ column density. 
As shown in \citet{pettini95}, strongly saturated \OI\ absorption lines can 
imply abundance ranges of up to $\sim$ 1000 without differences in the 
goodness of the profile fit. Examining the profile of the \OI\ line in 
Figure~\ref{figcap5}, it is clear that even this moderate-strength transition 
[log({\it f}$\cdot\lambda$) $=$ 0.974] is close to saturation in the central 
regions of the line. Furthermore, since these \fuse\ data probe multiple 
sightlines (see Figure~\ref{figcap1}), the line may be more heavily saturated 
than the profile shows, if a luminous continuum source is not shielded by the 
foreground \OI\ region. For these reasons, we more carefully explore the tests 
for, and potential effects of, \OI\ line saturation.

We can probe the amount of line saturation by subtracting the local 
rms value from the line profile and re-deriving the column density via the 
apparent optical depth method.  This exercise shows that the center of the 
line is within 1\,$\sigma$ of zero flux; this causes the inferred column 
density to grow quickly (see Equation~\ref{eq1} and discussion in 
\S~\ref{S3.2}).  This suggests that the \OI\ line is at least partially 
saturated.

The degree of this saturation can be tested by using tracer species to infer 
the column of neutral oxygen.  First, \SiII\ can be used as a proxy by 
assuming that the O/Si ratio is the same as the solar ratio (see {Lu \etal\ 
1998}\nocite{lu98} for details).  While uncertain, this method suggests that 
\OI\ saturation may cause the \OI\ column to be underestimated by as much as 
0.7 dex. Note, however, that Si can be depleted onto dust grains, and thus this
technique is not without error.  Second, \PII\ (which is less depleted onto 
dust grains) can be used as a tracer of the neutral oxygen abundance.  
Following \citet{lebouteiller04}, if [P/O] $=$ 0 is assumed for the neutral 
gas, then the implied column of \OI\ is nearly a full dex above that derived
from the actual \OI\ line profile.  Both of these proxy methods imply that the
\OI\ column has indeed been severely underestimated.  

Ideally, one would use other oxygen lines in the SiC channels to independently
test for line saturation.  As mentioned previously, these data lack the S/N
in the SiC channels to perform such a test.  Thus, in lieu of a large sample 
of O/Si and P/O studies of the neutral ISM of dwarf galaxies, for the 
remainder of this paper we explore the results of a column of \OI\ as derived 
above, log(N) $=$ 15.80\,$\pm$\,0.12, which implies an abundance offset when 
compared to the nebular regions.  We discuss the implications of this scenario
in \S~\ref{S4.3}.  We stress, however, that this \OI\ column density is a lower
limit and that the effects of line saturation may indeed be pronounced.

%-----------------------------------------------------------------------------%
\section{Discussion}
\label{S4}
%-----------------------------------------------------------------------------%

The most important results in this paper are as follows: 1) The discovery of 
outflowing coronal gas from a relatively low-luminosity dwarf starburst galaxy;
2) The discovery of diffuse H$_2$ absorption in a relatively distant, 
low-metallicity galaxy; 3) The discovery of a possible abundance offset 
between the nebular and neutral gas phases in a star-forming dwarf galaxy.  In
this section we discuss each of these points in more detail and compare the
values found for NGC\,625 with those found for other dwarf starbursts in the 
literature. 

%-----------------------------------------------------------------------------%
\subsection{Outflowing Coronal Gas}
\label{S4.1}
%-----------------------------------------------------------------------------%

The coronal gas content of low-mass systems that are undergoing outflow 
episodes is an important component of models of superbubble evolution. If 
the coronal gas density is high and radiative cooling is efficient, then 
outflow energies can be expended prior to the outflow breaking out of the disk
and venting hot gas, metals and energy into the surrounding IGM.  If, on the
other hand, the coronal gas density is low and the radiative cooling is 
inefficient, then the coronal gas does not radiate sufficient energy to slow or
stall galactic-scale outflows from low-mass galaxies.  While the evolution of 
outflows is complex and depends on various other factors, coronal gas cooling 
is one important effect that has only recently become accessible 
observationally.

In principal, by observing the absorption and emission profiles of outflowing 
\osix\ gas from a galaxy, one can obtain an estimate of the cooling rate of 
the gas and compare this to the estimated energy input into the ISM by the 
evolution of the massive stars associated with recent star formation (see
{Heckman \etal\ 2001}\nocite{heckman01a}).  However, such an estimate of the 
cooling rate in NGC\,625 is difficult for two reasons.  First, it is not 
straightforward to extrapolate to an unambiguous outflow cavity size estimate.
Second, the low outflow velocity of the \osix\ gas with respect to the 
molecular gas (and presumably the stellar population) in NGC\,625 ($\sim$ 30 
\kms) does not cleanly separate the absorption profile from the expected 
location of \osix\ emission (if we see emission only from the redshifted, back
side of a symmetric outflow along the line of sight). These factors complicate
estimates of the mass and density of the \osix\ gas, both of which are needed 
to constrain the cooling rate \citep[see][]{shull94}.

\fuse\ observations of \osix\ emission from other starbursting galaxies have 
shown that typical cooling rates in coronal gas are small.  \citet{heckman01a}
and \citet{hoopes03} estimate that coronal gas cooling is roughly comparable
to the energy radiated away in soft x-ray emission in the outflows in 
NGC\,1705 and M\,82.  Since the soft x-ray emission radiates only a small 
fraction of the input energy of the wind \citep{strickland00a}, the cooling 
rate via coronal gas appears to be small.  A similar column of \osix\ gas has 
been found in NGC\,625 (log(N(\osix)) = 14.32\,$\pm$\,0.08) and NGC\,1705 
(log(N(\osix)) = 14.26\,$\pm$\,0.08); because the outflow cavity is likely 
smaller in NGC\,625 than in NGC\,1705 (implying a smaller mass of cooling 
plasma), it is likely that radiative cooling via coronal gas is not a 
significant portion of the energy budget of NGC\,625.

Comparing to the literature, it appears that pronounced \osix\ absorption is
ubiquitous in actively star-forming systems with non-zero inclinations.  
Extensive observations through the Milky Way halo have shown an average \osix\ 
absorption column density of 14.38 \citep{savage03}. Similarly, 
\citet{howk02} and \citet{hoopes02} have shown that \osix\ absorption is 
present along tens of sightlines into the Large and Small Magellanic Clouds, 
with average column densities of 14.37 and 14.53, respectively.  Finally, 
studies of extragalactic dwarf systems with strong current star formation show 
at least weak \osix\ absorption when the major exciting clusters are observed 
directly (see {Heckman \etal\ 2001}\nocite{heckman01a} for NGC\,1705; 
{Lebouteiller \etal\ 2004}\nocite{lebouteiller04} for I\,Zw\,36; the present 
work for NGC\,625).  A large sample of \fuse\ dwarf galaxy observations 
spanning a range of galaxy properties (star formation rate, outflow velocities,
dust content, etc.) is needed to quantify the importance of \osix\ in the 
cooling portions of the outflows.

%-----------------------------------------------------------------------------%
\subsection{Diffuse Molecular Gas}
\label{S4.2}
%-----------------------------------------------------------------------------%

Diffuse H$_2$ is a ubiquitous component of the ISM of the Galaxy. Using 
{\it Copernicus} data, \citet{savage77} demonstrated a strong correlation 
between \HI\ column or reddening and the presence of diffuse H$_2$, with 
molecular gas usually found on sightlines with sufficient \HI\ to allow 
self-shielding or sufficient dust to allow the efficient formation of H$_2$.  
However, these early observations were limited to local sightlines due to 
instrumental limitations.  The higher sensitivity of {\it ORFEUS} allowed 
H$_2$ to be studied in the Magellanic Clouds (see, e.g., {de~Boer \etal\ 
1998}\nocite{deboer98}; {Richter \etal\ 1998}\nocite{richter98}). Most 
recently, \fuse\ has greatly expanded the amount of observational data 
available on diffuse H$_2$.  However, even though \fuse\ offers unprecedented 
spectral resolution and sensitivity (typical observations probe H$_2$ column
densities $\sim$ 10$^{15}$ cm$^{-2}$), there have remained few detections of 
H$_2$ in extragalactic environments.  Within the Local Group, diffuse H$_2$
has only been detected in the Milky Way (see, e.g., {Shull \etal\ 
2000}\nocite{shull00} and numerous other studies), the Magellanic Stream
\citep{richter01,sembach01}, the Magellanic Clouds \citep{tumlinson02}, and 
M\,33 \citep{bluhm03}.  As mentioned in \S~\ref{S1}, extragalactic detections 
of diffuse H$_2$ beyond the Local Group are even more rare.

Dwarf starburst galaxies such as NGC\,625 offer a unique opportunity to study
diffuse H$_2$ in the ISM, since they are typically UV-bright, are forming 
stars rapidly (implying a sizable molecular reservoir), and usually present 
relatively low extinctions due to their low metal abundances.  Even with these
apparent observational advantages, however, stringent column density limits 
of log(N(H$_2$)) $<$ 15 in the low-J levels are derived for the 
metal-deficient starburst galaxies I\,Zw\,18 \citep{vidalmadjar00}, NGC\,1705 
\citep{heckman01a}, Mrk\,59 {(Thuan \etal\ 2002)}\nocite{thuan02} and 
I\,Zw\,36 \citep{lebouteiller04}.  The new sample of \citet{hoopes04} also 
searches for diffuse H$_2$ gas in NGC\,3310, NGC\,4214, M\,83, and NGC\,5253; 
only M\,83 and NGC\,5253 show detectable H$_2$ gas in absorption against the 
background UV continuum of the massive stellar populations.  

\citet{hoopes04} interpret the low molecular fractions derived in their 
sample (typically {\it f}$_{\rm H_2}$ $\lsim$ 10$^{-5}$; i.e., similar to the 
fraction found in the present study for NGC\,625) to the stronger UV radiation
fields of these galaxies.  This stronger UV background will raise the minimum
values of extinction and \HI\ column necessary for diffuse H$_2$ to remain in 
the ISM without being destroyed.  They also note that some of the systems in 
their sample have been detected in CO tracer lines and that the inferred
molecular masses exceed the derived diffuse H$_2$ masses by many orders of 
magnitude.  

The detection of H$_2$ absorption in NGC\,625 highlights the importance of 
ISM geometry in the interpretation of these types of observations.  As shown 
by \citet{cannon02}, the dust distribution (and hence potentially productive 
areas of H$_2$ formation) does not always follow the distribution of UV light.
This implies that UV absorption experiments for diffuse H$_2$ may not sample 
the sightlines expected to show the highest columns of molecular gas.  In the
case of NGC\,625, we have a combination of sightlines into a moderately
dusty starburst region (extinctions toward the massive clusters of up to $\sim$
1 magnitude in the V-band; see {Cannon \etal\ 2003}\nocite{cannon03}) that 
has a moderate UV radiation field (since no higher-level H$_2$ transitions are
detected).  Here, the potential for detection of H$_2$ may be optimized, given
the correlation between dust and H$_2$ \citep[see, e.g.,][]{savage77}.  

The simplest interpretation of these H$_2$ observations of nearby dwarf 
starburst and low-metallicity galaxies appears to be one where the covering 
factor of diffuse H$_2$ clouds is low.  This implies that most of the 
molecular material is confined to relatively small, dense molecular clouds,
as predicted in models of the low-metallicity ISM \citep[see, 
e.g.,][]{maloney88,norman97,pak98,bolatto99}.  Since the strength of the UV 
radiation field appears to be an important factor both theoretically and 
observationally, it may be expected that sightlines to the UV-brightest 
clusters may not show diffuse H$_2$ in absorption.  The complicated geometry 
of dust, neutral and molecular gas, and the current stellar populations will 
require a much larger sample to fully understand the nature of diffuse H$_2$ 
absorption in extragalactic environments.

%-----------------------------------------------------------------------------%
\subsection{Abundances in the Neutral and Nebular Gas}
\label{S4.3}
%-----------------------------------------------------------------------------%
 
In these data we find an apparent abundance difference between the neutral and
nebular gas regions in NGC\,625.  Interestingly, the N/O ratio is, within 
errors, identical in the neutral and nebular gas phases. We again emphasize 
that our \ion{O}{1} $\lambda$\,1039.230 \AA\ line may be saturated (hence 
reducing the offset between N and O abundances), and that our absolute 
abundances are comparatively uncertain due to the large error on our measured 
\HI\ column within NGC\,625.  However, taken at face value, the offset is large
enough ($\sim$ 0.9 dex) to warrant a more thorough discussion of these values 
in other systems and the potential implications for the evolution of low-mass 
galaxies. 

As shown in Table~\ref{t3}, it appears that such abundance offsets between 
nebular and neutral gas are common in strongly star-forming dwarf galaxies 
studied to date with \fuse.  Abundance differences of $>$ 0.5 dex have been
found in NGC\,1705 \citep{heckman01a}, I\,Zw\,18 ({Aloisi \etal\ 
2003}\nocite{aloisi03}; but see also {Lecavelier des Etangs \etal\ 
2004}\nocite{lecavelierdesetangs04} for an alternative treatment), I\,Zw\,36
\citep{lebouteiller04}, and Mrk\,59 \citep{thuan02}.  These abundance 
differences may be caused by \fuse\ sightlines sampling two different 
components of the ISM of dwarf galaxies:  lower-abundance halo gas, and 
higher-abundance gas nearer to the active star formation regions.  

\begin{figure}
\plotone{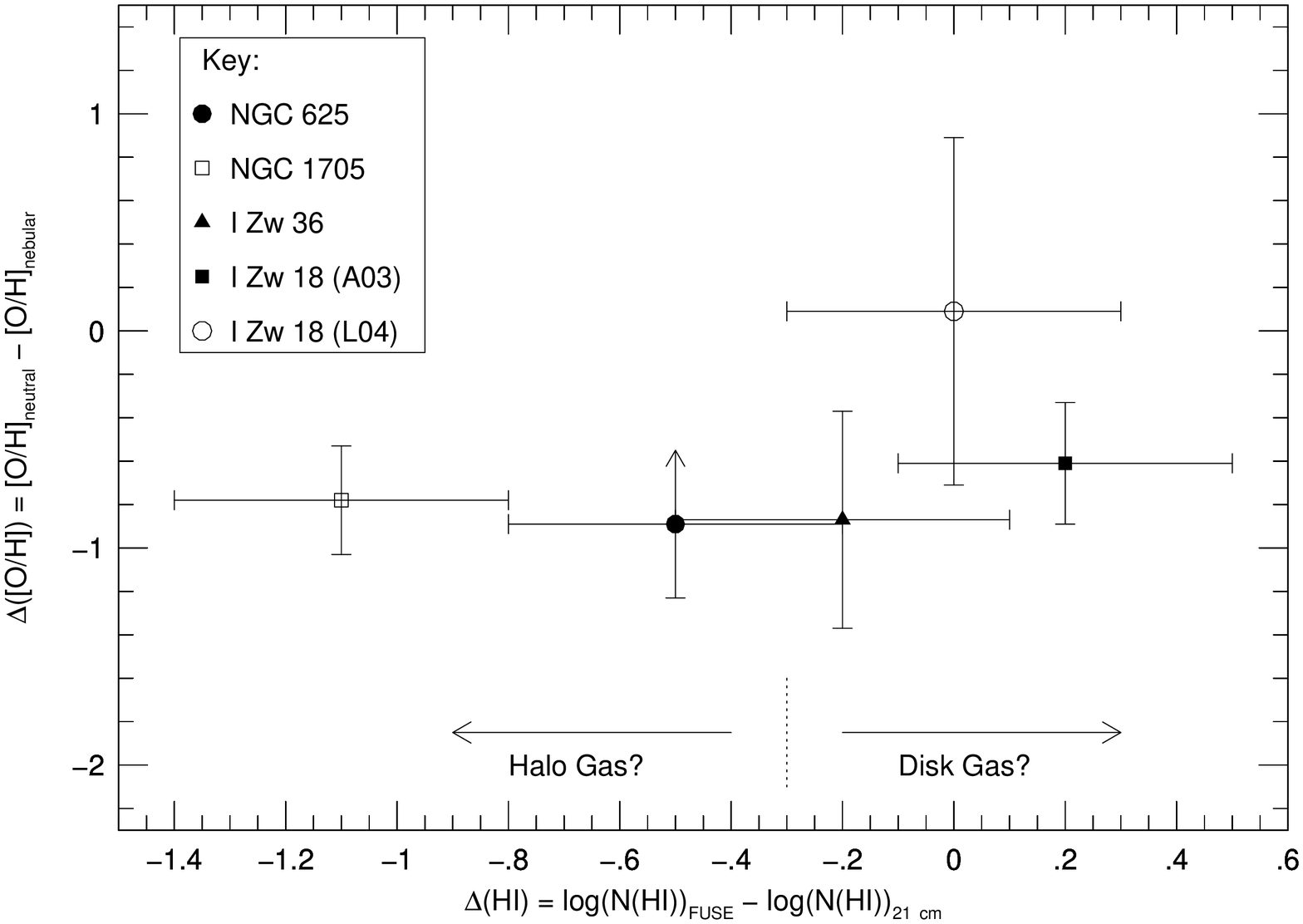}
\caption{Differences between neutral and nebular gas abundances plotted versus
the difference between \HI\ absorption column density as probed by \fuse\ 
observations and \HI\ 21\,cm emission column density, for 4 different dwarf 
starburst galaxies.  In the key, A03 corresponds to the study of 
\citet{aloisi03}, and L04 corresponds to the study of 
\citet{lecavelierdesetangs04}.  Errorbars on $\Delta$(\HI) are difficult to 
estimate, and are shown at the 0.3 dex level.  We find no obvious correlation 
between these two values; rather, abundance offsets between the neutral and 
nebular gas appear to be common in the dwarf galaxies shown here.  To the left
of the dotted vertical line at $\Delta$(\HI) $= -$0.3, one expects to probe 
``halo'' gas via absorption line spectroscopy; to the right of this line, one
expects only to probe gas close to the background sources within the disk.}
\label{figcap7}
\end{figure}

A simple test of this ``disk vs. halo'' scenario can be performed by comparing 
the size of the abundance difference with the difference between the \HI\ gas 
column seen in emission and the \HI\ column that can only be foreground to the
UV-luminous regions (i.e., the \HI\ gas probed in \fuse\ observations).  Our 
reasoning is simple: if the absorption arises in the disk of the galaxy, the
column density should be comparable to the disk column density, but if the
absorption arises in a halo, the column density should be smaller. We show in 
Figure~\ref{figcap7} a plot of the size of the offset between neutral and 
nebular gas ($\Delta$($[$O/H$]$) $=$ $[$O/H$]$$_{\rm neutral}$ $-$ 
$[$O/H$]$$_{\rm nebular}$) and the difference in \HI\ columns as seen in 
absorption and in emission toward the same region ($\Delta$(\HI) $=$ 
log(N(\HI))$_{\rm FUSE}$ $-$ log(N(\HI))$_{\rm 21\,cm}$).  Depending on the 
value chosen for I\,Zw\,18, there could be a trend of increasing
$\Delta$($[$O/H$]$) with increasing $\Delta$(\HI). It appears that each
system under study (with the possible exception of I\,Zw\,18) shows some value 
of abundance offset between the nebular and neutral gas (see notes to 
Table~\ref{t3} for more details).  Each of these investigations is susceptible
to various sources of error that will require a larger sample to overcome; 
however, if these results are supported by further data, it appears that many 
dwarf galaxies may have a halo of lower-metallicity gas that surrounds the 
actively star-forming regions as probed by emission line spectroscopy.

This result has important implications for the chemical evolution of 
star-forming dwarf galaxies, since it implies that widespread enrichment 
episodes have preceded the current bursts that dominate the bolometric
luminosity of these systems.  Previous works have postulated a prompt
``self-enrichment'' scenario in the nebular gas region that would be evidenced 
by localized enrichment near current \HII\ regions \citep[e.g.,][]{kunth86}.
This effect has been shown to be small, since local abundance variations 
are not seen in the nebular regions surrounding massive starbursts in low-mass
galaxies (see {Kobulnicky \& Skillman 1996}\nocite{kobulnicky96}, {Kobulnicky 
\& Skillman 1997}\nocite{kobulnicky97b}, {Kobulnicky \etal\ 
1997}\nocite{kobulnicky97a}, {Kobulnicky \& Skillman 
1998}\nocite{kobulnicky98}, {Legrand \etal\ 2000}\nocite{legrand00b}, and 
references therein).

The present offset between neutral and nebular gas abundances would require a 
different enrichment scenario than ``localized enrichment'', since the 
absorbing columns are, by definition, foreground to the starburst regions.  
A potential scenario that would be consistent with these data is one where 
low-level star formation persists in dwarf galaxies for extended periods of 
time, allowing the oxygen and nitrogen abundances to be elevated in the ISM 
and efficiently mixed with the surrounding neutral gas.  This low-level star
formation rate would elevate the neutral gas abundances above the primordial 
value, and the higher nebular values could then be produced by more recent star
formation in the current major star formation regions of these systems. The 
close agreement in N/O between the nebular and neutral gas also points toward
central creation and geometrical dilution over a long time period; otherwise 
the different production timescales for N and O become problematic. A larger 
\fuse\ sample of actively star forming dwarf galaxies would be most beneficial 
in addressing the abundance offsets between nebular and neutral gas phases.

If the absorption spectra of dwarf galaxies obtained by {\it FUSE} are indeed 
sampling a halo of neutral gas and are not primarily produced in the disks, 
then they provide a very important probe of a virtually unstudied component of
the dwarf galaxy ISM. \citet{kennicutt01} have emphasized the importance of 
measuring abundances in the neutral gas in the outer parts of dwarf galaxies. 
In spiral galaxies, it is well known that there are chemical abundance 
gradients in the sense of lower abundances in the outer parts of the systems.
However, in dwarf galaxies it is generally assumed that the entire \HI\ disk 
has the same metallicity as measured in the \HII\ regions, and this is a very 
uncertain assumption.  The physical basis for this assumption is the general 
uniformity of \HII\ region abundances in dwarf galaxies ({Kobulnicky \& 
Skillman 1997}\nocite{kobulnicky97b} and references therein), and the 
inference that the whole \HI\ disk is kept at a rather uniform chemical 
abundance by the rapid transportation of the metals in a hot phase of the ISM 
\citep[][]{clayton93,tenoriotagle96}.  However, the \HII\ regions only sample 
the inner parts of the \HI\ disk. In some dwarf galaxies, as much as 90\% of 
the neutral hydrogen lies outside of the Holmberg radius (e.g., DDO\,154; 
{Carignan \& Freeman 1988}\nocite{carignan88}; {Carignan \& Purton 
1998}\nocite{carignan98}). If dwarf galaxies do have chemical abundance 
gradients, then assuming that the chemical abundances are constant 
overestimates the total metal content of the galaxy, and leads to a  
misinterpretation of their evolutionary status (e.g., artificially inflating 
the calculated effective yield). The edge-on orientation of NGC\,625 implies 
that at least some of the absorption is occurring in the outer parts of the 
galaxy and thus providing a probe of this relatively unexplored ISM component. 

The extended mission of \fuse\ offers an ideal opportunity to test for this
important evolutionary scenario.  One would ideally seek a sample of luminous, 
metal-poor (\lsim\ 10\% Z$_{\odot}$) dwarf systems that have low intrinsic 
\HI\ foreground columns, low foreground and internal extinctions, along 
high-visibility sightlines not contaminated by intermediate- or high-velocity 
clouds.  Deep integrations on such targets will provide sufficient S/N to allow
inter-comparison of the columns derived from oxygen lines with different 
oscillator strengths throughout the \fuse\ spectral region.  With the effects 
of line saturation eliminated, such a sample would allow the exploration of 
this potentially important ISM phase in dwarf galaxies.   

%-----------------------------------------------------------------------------%
\section{Conclusions}
\label{S5}
%-----------------------------------------------------------------------------%

We have presented new \fuse\ spectroscopy of the dwarf starburst galaxy 
NGC\,625.  These data allow a detailed investigation of multiple phases of the 
ISM of the galaxy, including the molecular, neutral, ionized and coronal gas 
contents.  We use these data to study the kinematics of the ISM, the diffuse 
H$_2$ content, and the abundances in the neutral gas phase. 

Our first major result is that \osix\ absorption has been detected in a 
relatively low-velocity outflow.  This detection adds to the sample of 
extragalactic systems showing outflowing coronal gas.  With these data alone 
we cannot constrain the efficiency of cooling in the coronal gas; however, 
comparing to the literature and given the strength of the \osix\ absorption 
feature, we suggest that radiative cooling is likely not a dominant mechanism 
in the loss of energy from the NGC\,625 outflow. 

Our second major result is that we have detected low rotational level 
transitions from the Lyman and Werner bands of diffuse H$_2$ gas in NGC\,625.
This is one of only a few dwarf galaxies outside of the Local Group to show 
diffuse H$_2$ in \fuse\ observations. It is likely that the geometry of the 
stellar populations, dust, neutral and dense molecular gas all play a role in 
determining if H$_2$ is detectable in such systems.  We interpret the low 
molecular fraction, which is similar to that found in other dwarf starburst 
and low-metallicity galaxies from \fuse\ studies, to the low covering factor 
of H$_2$ clouds, which are expected to be comparatively small and dense in 
these actively star-forming, metal-deficient environments. 

Our final important result is that we have found a potential abundance offset 
between the nebular and neutral gas phases of NGC\,625.  While the absolute 
oxygen abundance is hindered by potential saturation and the intrinsic \HI\ 
column is only constrained to $\pm$\,0.3 dex, the magnitude of the offset 
($\sim$ 0.9 dex) suggests that it is likely real.  Interestingly, the N/O ratio
is found to be the same in both the nebular and neutral phases.  Similar 
results have been found for other dwarf galaxies studied with \fuse, suggesting
that dwarf galaxies may in general host a lower-metallicity halo of neutral gas
that can only be probed by absorption line spectroscopy.  A larger sample of 
systems will help to shed light on this important evolutionary scenario.

%-----------------------------------------------------------------------------%
\acknowledgements
%-----------------------------------------------------------------------------%

We thank Max Pettini for several valuable comments including insights into 
aspects of oxygen line saturation; S.~R. McCandliss for making the {\it 
H$_2$ools} package available; and Henry Lee \& Simon Strasser for helpful 
discussions. We are also grateful to the anonymous referee for a careful 
reading of this manuscript and numerous helpful comments.  J.\,M.\,C. was 
supported by NASA Graduate Student Researchers Program (GSRP) Fellowship NGT 
5-50346. E.\,D.\,S. is grateful for partial support from NASA LTSARP grant 
NAG5-9221 and the University of Minnesota.  This research has been supported 
by NASA grant NNG 04GD25G. This research has made use of the NASA/IPAC 
Extragalactic Database (NED) which is operated by the Jet Propulsion 
Laboratory, California Institute of Technology, under contract with the 
National Aeronautics and Space Administration, and NASA's Astrophysics Data 
System. 

%-----------------------------------------------------------------------------%
\clearpage
%\bibliographystyle{apj}                                                 
%\bibliography{references,unread}   

%-----------------------------------------------------------------------------%

\clearpage

\clearpage

\clearpage

\clearpage

\clearpage

\clearpage

\clearpage

\clearpage
\begin{deluxetable}{ccc}
\tabletypesize{\scriptsize}
\tablecaption{Basic Parameters of NGC\,625}
\tablewidth{0pt}
\tablehead{
\colhead{Property}         
&\colhead{Value} 
&\colhead{Reference}}
\startdata
Distance (Mpc)  	&3.89\,$\pm$\,0.22	&\citet{cannon03}\\
M$_B$			&$-$16.3		&\citet{marlowe97}\\
{\it b, l}\,\tablenotemark{a}	&273.7\degree, $-$73.1\degree	&$--$\\
E(B$-$V)		&0.016			&\citet{schlegel98}\\
12\,$+$\,log(O/H) 	&8.14\,$\pm$\,0.02	&\citet{skillman03b}\\
Current SFR (\msun\,yr$^{-1}$)  &0.05  		&\citet{skillman03a}\\
\HI\ Mass (10$^8$ \msun) &1.1\,$\pm$\,0.1	&\citet{cannon04a}\\
V$_{Helio}$ (\kms)	&413\,$\pm$\,5 		&\citet{cannon04a}\\
\enddata
\tablenotetext{a}{Galactic longitude and latitude.}
\label{t1}
\end{deluxetable}

\clearpage
\begin{deluxetable}{ccccccc}
\tabletypesize{\scriptsize}
\tablecaption{Important Line Parameters}
\tablewidth{0pt}
\tablehead{
\colhead{Line}         
&\colhead{$\lambda_{\rm 0}$\tablenotemark{a}} 
&\colhead{$\lambda_{\rm NGC\,625}$}
&\colhead{log(f$\cdot\lambda$)\tablenotemark{a}}
&\colhead{$V_{\rm NGC\,625}$}
&\colhead{W$_{\rm NGC\,625}$\tablenotemark{b}}
&\colhead{log(N)\tablenotemark{c}}\\
\colhead{I.D.}         
&\colhead{(\AA)} 
&\colhead{(\AA)}
&\colhead{}
&\colhead{(\kms)}
&\colhead{(\AA)}
&\colhead{}}
\startdata
H$_2$ L R(0)(2-0)\tablenotemark{d} &1077.140  &1078.60\,$\pm$\,0.02 &1.099 
&406\,$\pm$\,6 &0.07\,$\pm$\,0.007 &14.92\,$\pm$\,0.19\\
H$_2$ L R(1)(4-0)\tablenotemark{d} &1049.960  &1051.38\,$\pm$\,0.03 &1.213 
&406\,$\pm$\,9 &0.12\,$\pm$\,0.01 &15.01\,$\pm$\,0.15\\
H$_2$ L R(2)(5-0)\tablenotemark{d} &1038.689  &1040.07\,$\pm$\,0.02 &1.234 
&399\,$\pm$\,6 &0.04\,$\pm$\,0.004 &14.34\,$\pm$\,0.19\\
H$_2$ L P(1)(5-0)\tablenotemark{d} &1038.157  &1039.51\,$\pm$\,0.02 &0.954 
&391\,$\pm$\,6 &0.05\,$\pm$\,0.005 &14.82\,$\pm$\,0.18\\
H$_2$ W R(2)(0-0)\tablenotemark{d} &1009.024  &1010.36\,$\pm$\,0.02 &1.198 
&397\,$\pm$\,6 &0.05\,$\pm$\,0.005 &14.59\,$\pm$\,0.18\\
H$_2$ L R(1)(8-0)\tablenotemark{d} &1002.449  &1003.80\,$\pm$\,0.03 &1.262 
&404\,$\pm$\,9 &0.09\,$\pm$\,0.009 &14.87\,$\pm$\,0.15\\
	
%H Ly$\beta$ &1025.7218  &--               &1.733 &--        &--   &--    &1\\

%\CII        &1036.3367  &--               &2.088 &--        &--   &--    &2\\
\CII$^*$    &1037.018  &1038.30\,$\pm$\,0.02 &2.088 &371\,$\pm$\,6 &--   &--\\
     	
\NI         &1134.980  &1136.43\,$\pm$\,0.02 &1.674 &383\,$\pm$\,6 
&0.10\,$\pm$\,0.01 &14.63\,$\pm$\,0.12\\	
\NI         &1134.415  &1135.88\,$\pm$\,0.02 &1.512 &387\,$\pm$\,6 
&0.09\,$\pm$\,0.009 &$<$ 14.80\,$\pm$\,0.12\\
%\NI         &1134.1653  &1136.43\,$\pm$\,0.03 &1.219 &395\,$\pm$\,9 &0.04 
%&14.39 &--\\
%\NII        &1083.9937  &--               &2.079 &--  &--   &--    &1\\
 
\OI         &1039.230  &1040.55\,$\pm$\,0.02 &0.974 &381\,$\pm$\,6 
&0.24\,$\pm$\,0.02 &$\ge$ 15.80\,$\pm$\,0.12\tablenotemark{e}\\
\OVI        &1031.926  &1033.20\,$\pm$\,0.02 &2.136 &370\,$\pm$\,6 
&0.20\,$\pm$\,0.02 &14.32\,$\pm$\,0.08\\
     	
\SiII       &1020.699  &1022.03\,$\pm$\,0.03 &1.234 &391\,$\pm$\,9 
&0.19\,$\pm$\,0.02 &15.37\,$\pm$\,0.15\\
      	
\PII        &1152.818  &1154.31\,$\pm$\,0.02 &2.451 &388\,$\pm$\,6 
&0.09\,$\pm$\,0.009 &13.54\,$\pm$\,0.16\\

\SIII       &1012.495  &1013.79\,$\pm$\,0.02 &1.647 &383\,$\pm$\,6 
&0.38\,$\pm$\,0.04 &15.04\,$\pm$\,0.15\\
     	
\ArI        &1066.660  &1068.10\,$\pm$\,0.02 &1.857 &405\,$\pm$\,6 
&0.06\,$\pm$\,0.006 &14.20\,$\pm$\,0.20\\
\ArI        &1048.220  &1049.58\,$\pm$\,0.02 &2.440 &389\,$\pm$\,6 
&0.12\,$\pm$\,0.01 &13.93\,$\pm$\,0.10\\
     	
\FeII       &1144.938  &1146.45\,$\pm$\,0.02 &1.978 &396\,$\pm$\,6 
&0.28\,$\pm$\,0.03 &14.72\,$\pm$\,0.16\\
%\FeII	    &1143.2260  &--               &1.342 &--  &--   &--    &2\\
%\FeII	    &1142.3656  &1143.86\,$\pm$\,0.04 &0.661 &392\,$\pm$\,12 &0.07 
%&15.09 &1, 3\\
%\FeII	    &1133.6654  &--               &0.728 &--  &--   &--    &1, 3\\
%\FeII	    &1127.0984  &--               &0.102 &--  &--   &--    &1\\
%\FeII	    &1125.4477  &--               &1.244 &--  &--   &--    &1\\
%\FeII	    &1121.9748  &1123.43\,$\pm$\,0.03 &1.512 &389\,$\pm$\,9 &--   
%&--    &4\\
%\FeII	    &1112.0480  &1113.54\,$\pm$\,0.03 &0.695 &402\,$\pm$\,9 &0.04 
%&14.55 &1\\
%\FeII	    &1096.8769  &1098.29\,$\pm$\,0.02 &1.554 &388\,$\pm$\,6 &0.18 
%&14.72 &--\\
%\FeII	    &1063.9718  &--               &0.704 &--  &--   &--    &1\\
\FeII	    &1063.176  &1064.61\,$\pm$\,0.02 &1.765 &404\,$\pm$\,6 
&0.17\,$\pm$\,0.02 &14.73\,$\pm$\,0.15\\
%\FeII	    &1055.2617  &1056.68\,$\pm$\,0.03 &0.812 &403\,$\pm$\,9 &0.05 
%&13.76 &1\\ 
%\FeIII      &1122.5240  &--               &1.786 &--  &--   &--    &4\\ 
\enddata\\
%Table Notes: 1 - Low S/N feature.\\
%2 - Line blended with Galactic absorption feature.\\
%3 - Complicated line structure due to neighboring absorption feature.\\
%4 - Line blended with NGC\,625 absorption feature.\\
\tablenotetext{a}{All atomic data are taken from \citet{morton03}.  All 
molecular data is taken from the {\it H$_2$ools} website; see 
http://www.pha.jhu.edu/~stephan/h2ools2.html and \citet{mccandliss03}.}
\tablenotetext{b}{Representative errors on the equivalent width values are 
$\pm$\,10\%.}
\tablenotetext{c}{Column densities derived from the apparent optical depth 
method; see \citet{savage91}.}
\tablenotetext{d}{The labels ``L'' and ``W'' denote Lyman or Werner bands,
respectively.}
\tablenotetext{e}{The \OI\ $\lambda$\,1039.230 \AA\ line may be saturated; 
without a second oxygen line for comparison, there is no empirical way to 
test the saturation of this line.  We note this uncertainty and discuss it
further in \S~\ref{S3.3.3}.}
\label{t2}
\end{deluxetable}

\clearpage
\begin{deluxetable}{ccccccccc}
\rotate
\tabletypesize{\scriptsize}
\tablecaption{Neutral vs. Nebular Abundances in Low-Metallicity 
Galaxies\tablenotemark{a}}
\tablewidth{0pt}
\tablehead{
\colhead{Galaxy} 
&\colhead{Nebular}                    
&\colhead{Neutral} 
&\colhead{Nebular}                    
&\colhead{Neutral}                    
&\colhead{Nebular}
&\colhead{Neutral}
&\colhead{$\Delta$(\HI)}
&\colhead{Reference(s)}\\
&\colhead{$[$N/H$]$\tablenotemark{b}} 
&\colhead{$[$N/H$]$\tablenotemark{c}} 
&\colhead{$[$O/H$]$\tablenotemark{b}} 
&\colhead{$[$O/H$]$\tablenotemark{c}} 
&\colhead{log(N/O)\tablenotemark{b}} 
&\colhead{log(N/O)\tablenotemark{c}} 
&\colhead{(dex)\tablenotemark{d}} &}
\startdata
NGC\,625	
&$-$1.07\,$\pm$\,0.13	
&$-$1.80\,$\pm$\,0.34	
&$-$0.47\,$\pm$\,0.07		
&$\ge -$1.36\,$\pm$\,0.33	
&$-$1.33\,$\pm$\,0.02	
&$\le -$1.17\,$\pm$\,0.15	
&$< -$\,0.5	
&1, 2, 3\\
NGC\,1705	
&$-$1.47\,$\pm$\,0.14
&$-$2.16\,$\pm$\,0.24
&$-$0.45\,$\pm$\,0.07	
&$-$1.23\,$\pm$\,0.22		
&$-$1.75\,$\pm$\,0.06	
&$-$1.66\,$\pm$\,0.11	
&$-$\,1.1	
&4, 5, 6\\
I\,Zw\,18\tablenotemark{e}	
&$-$2.31\,$\pm$\,0.13	
&$-$2.88\,$\pm$\,0.11	
&$-$1.45\,$\pm$\,0.05	
&$-$2.06\,$\pm$\,0.28	
&$-$1.59\,$\pm$\,0.05	
&$-$1.54\,$\pm$\,0.26	
&$< +$\,0.2 	
&7, 8, 9\\
I\,Zw\,18\tablenotemark{f}	
&$-$2.31\,$\pm$\,0.13	
&$-$3.00\,$\pm$\,0.18	
&$-$1.45\,$\pm$\,0.05	
&$-$1.36\,$\pm$\,0.80	
&$-$1.59\,$\pm$\,0.05	
&$-$2.40\,$^{+\,0.6}_{-\,0.8}$	
&$< +$\,0.0 	
&7, 10, 9\\
I\,Zw\,36	
&$-$1.65\,$\pm$\,0.11	
&$-$2.81\,$\pm$\,0.38	
&$-$0.89\,$\pm$\,0.05	
&$-$1.76\,$\pm$\,0.50\tablenotemark{g}	
&$-$1.49\,$\pm$\,0.01	
&$-$1.80\,$\pm$\,0.60\tablenotemark{g}	
&$-$\,0.2\tablenotemark{h}	
&11, 12, 13\\
Mrk\,59 	
&$-$1.46\,$\pm$\,0.11	
&$-$2.08\,$\pm$\,0.59	
&$-$0.67\,$\pm$\,0.05	
&$-$1.66\,$\pm$\,0.30
&$-$1.52\,$\pm$\,0.01	
&$-$1.15\,$\pm$\,0.5	
&N/A\tablenotemark{i}
&11, 14\\
\enddata
\vspace{-0.5 cm}
\tablerefs{1 - {Skillman \etal\ 2003b}\nocite{skillman03b}; 2 - This work; 3 -
{Cannon \etal\ 2004a}\nocite{cannon04a}; 4 - {Lee \& Skillman 
2004}\nocite{lee04}; 5 - {Heckman \etal\ 2001}\nocite{heckman01a}; 6 - {Meurer
\etal\ 1998}\nocite{meurer98}; 7 - {Skillman \& Kennicutt 
1993}\nocite{skillman93}; 8 - {Aloisi \etal\ 2003}\nocite{aloisi03}; 9 - 
{van~Zee \etal\ 1998}\nocite{vanzee98b}; 10 - {Lecavelier des Etangs \etal\ 
2004}\nocite{lecavelierdesetangs04}; 11 - {Izotov \& Thuan 
1999}\nocite{izotov99a}; 12 - {Lebouteiller \etal\ 
2004}\nocite{lebouteiller04}; 13 - {Stil \& Israel 2002}\nocite{stil02b}; 14 - 
{Thuan \etal\ 2002}\nocite{thuan02}}
\tablenotetext{a}{The abundances here use the solar nitrogen value from 
\citet{holweger01} and the solar oxygen abundance from \citet{asplund04}.}
\tablenotetext{b}{Derived from nebular spectroscopy of the most luminous \HII\ 
regions in the galaxies, since these will be most heavily weighted in UV \fuse\
measurements.}
\tablenotetext{c}{Derived from \fuse\ spectroscopy.}
\tablenotetext{d}{Defined as the difference between the \HI\ absorption column 
density toward the heavily-weighted UV sightlines as probed by \fuse, and the 
\HI\ 21\,cm emission column density (for the area nearest to the \fuse\ 
aperture): $\Delta$(\HI) $=$ log(N(\HI))$_{\rm FUSE}$ $-$ 
log(N(\HI))$_{\rm 21\,cm}$.}
\tablenotetext{e}{Neutral gas values from analysis of \fuse\ data by 
\citet{aloisi03}.} 
\tablenotetext{f}{Neutral gas values from analysis of \fuse\ data by 
\citet{lecavelierdesetangs04}.} 
\tablenotetext{g}{Calculated using Phosphorus as a tracer of oxygen abundance;
see discussion in \citet{lebouteiller04}.}
\tablenotetext{h}{Only peak \HI\ 21\,cm emission column density value is 
available; see \citet{stil02b}.}
\tablenotetext{i}{Detailed \HI\ emission observations of Mrk\,59 are not yet 
available in the literature.}
\label{t3}
\end{deluxetable}
\end{document}